\documentclass[a4,11pt]{article} 
\usepackage{amssymb}
\usepackage{amsmath}
\usepackage[pdftex]{graphicx}
\usepackage{cite}
\usepackage{slashed}
\usepackage{bm}
\usepackage{fancyhdr}
\usepackage{float}
\usepackage{tocloft}
\usepackage[top=2.5cm, bottom=2.0cm, left=2.5cm, right=2cm]{geometry}

\begin{document}

\begin{center}
{\large \bf Investigation of the scalar unparticle and anomalous couplings at muon colliders in final states with  multiple photons in the Randall- Sundrum model}\\
\vspace*{1cm}

 { Bui Thi Ha Giang$^{a,}$ \footnote{ Coressponding authors: giangbth@hnue.edu.vn; soadangvan@gmail.com.vn }}, {Dang Van Soa $^{b,}$ , Le Mai Dung$^{a}$}\\

\vspace*{0.5cm}
 $^a$ Hanoi National University of Education, 136 Xuan Thuy, Hanoi, Vietnam\\
 $^b$ Faculty of Applied Sciences, University of Economics - Technology for Industries,\\
  456 Minh Khai, Hai Ba Trung, Hanoi, Vietnam
\end{center}

\begin{abstract}
\hspace*{0.5cm} The influence of the scalar  unparticle and  anomalous couplings at muon colliders in final states with  multiple photons in the Randall-Sundrum model is evaluated in detail. The results indicate that with fixed collision  energies, the total cross-sections for the production of multiple photons depend strongly on the polarization of the muon beams, the parameters of unparticle physics (the scaling dimension $d_{U}$,  operator $\mathcal{O}_{U}$, the energy scale $\Lambda_{U}$) and also the strength of  anomalous couplings. Numerical evaluation shows that the cross-sections for the production of four photonn in finale states with the contribution of scalar anomalous couplings are much larger than that of the unparticle under the same conditions. In the Higgs-radion mixing, the cross sections achieve the maximum value at the radion-dominated state, $m_{\phi} = 125$ GeV, in which the cross-section is much enhanced and can be  measurable  in current experiments.\\
\end{abstract}

\section{Introduction}
\hspace*{1cm}The Standard model (SM) is the successful model in describing the elementary particle picture. Recently, Higgs signal at 125 GeV is discovered by the ATLAS and CMS collaborations \cite{Aad, Chat}, which has completed the particle spectrum of the SM. Although the SM has been considered to be successful model, the model suffers from many theoretical drawbacks. In 1999, Lisa Randall and Raman Sundrum suggested the Randall-Sundrum (RS) model to extend the SM and solve the hierarchy problem naturally \cite{Ran}. The RS setup involves two three-branes bounding a slice of 5D compact anti-de Sitter space. Gravity is localized UV brane, while the SM fields are supposed to be localized IR brane. The separation between the two 3-branes leads directly to the existence of an additional scalar called the radion ($\phi$ ), corresponding to the quantum fluctuations of the distance between the two 3-branes \cite{Fra}. Phenomenology of scalar particles in RS model has been intensively studied in \cite{csa, dav, dominici, soa, soa1, ali, bha, boo, soael, ahm}. \\
 \hspace*{1cm}At TeV scale, the scale invariant sector has been considered as an effective theory and that if it exists, it is made of unparticle suggested by Geogri \cite{georgi,georgi2}. Based on the Banks-Zaks theory \cite{banks}, unparticle stuff with nontrivial scaling dimension is considered to exist in our world and this opens a window to test the effects of the possible scalar invariant sector, experimentally \cite{chenhe}.
 The effects of unparticle on properties of high energy colliders have been intensively studied in Refs.\cite{dong,pra,alan,maj,kuma,sahi,kiku,chen,kha,fried,giang,giang1}. The discovery of unparticle could have far-reaching implications for
particle physics and cosmology \cite{beze, lee, arty, muonE, putten, wu}. Bounds from LEP on unparticle interactions with electroweak bosons are investigated in Ref.\cite{scott}. Moreover, the multiphoton signals at LHC through the unparticle self-interaction have been studied in detail in Ref.\cite{alie}. Recently, the possibility of the unparticle has been studied with CMS detector at the LHC \cite{cms15,cms16,cms17}.\\
\hspace*{1cm} In  the high energy collisions, new scalar particles produced can decay directly into the SM light particles, so the signature of multiple leptons/photons in final states is considered as one of the golden channels to discover new physics at the LHC. Searchs for new physics in final states with multiple leptons in high energy collisions are considered in detail\cite{alb}. Recently, the search for the exotic decay of the Higgs boson into two light pseudoscalars with four photons in  final states at the high energy colliders has been presented\cite{arm}. In this work, we will evaluate the influence of scalar unparticle and scalar anomalous couplings on production of multiple photons in final states at multi TeV- muon colliders in the Randall-Sundrum model. The layout of this paper is as follows. In Section II, we introduce the mixing of Higgs-radion and the scalar unparticle in the RS model. The influence of the scalar unparticle and anomalous couplings on the production of mutiple photons  is given in detail in Sections III. Finally, we summarize our results and make conclusions in Section IV.
 
\section{The mixing of Higgs - radion and the scalar unparticle in Randall-Sundrum model}
\subsection{The mixing of Higgs - radion}
\hspace*{1cm}The RS model is based on a 5D spacetime with two three-branes: the UV brane and the IR brane. All the SM and dark matter (DM) fields, excepted for gravity, are assumed at IR brane. Gravity lives on the second three-branes. The RS model consider a non-factorizable 5-dimensional metric in the form 
\begin{equation}
d_{s}^{2} = e^{-2 \sigma} \eta_{\mu \nu} d x^{\mu} d x^{\nu} - r_{c}^{2} d y^{2},
\end{equation}
where $\sigma = k r_{c}|y|$, $k$ is the curvature along the 5th-dimension, $r_{c}$ is the length-scale which is related to the size of the extra-dimension. For $\sigma=k r_c \sim 10$ the RS scenario can address the hierarchy problem. The action in $5 \mathrm{D}$ is:
\begin{equation}
S=S_{\text {gravity }}+S_{\mathrm{IR}}+S_{\mathrm{UV}}.
\end{equation}
where $S_{\mathrm{IR}}, S_{\mathrm{UV}}$ are the brane actions for the two branes.\\
\hspace*{1cm} The possibility of mixing between gravity and the electroweak sector can be described by the following action
\begin{equation}
S_{\xi } =\xi \int d^{4}x \sqrt{g_{vis} } R(g_{vis} )\hat{H}^{+} \hat{H},
\end{equation}
where $\xi $ is the mixing parameter \cite{ahm, ali}, $R(g_{vis})$ is the Ricci scalar for the metric $g_{vis}^{\mu \nu } =\Omega _{b}^{2} (x)(\eta ^{\mu \nu } +\varepsilon h^{\mu \nu } )$ induced on the visible brane, $\Omega _{b} (x) = e^{-kr_{c} \pi} (1 + \frac{\phi_{0}}{\Lambda _{\phi }})$ is the warp factor, $\Lambda _{\phi }$ is the vacuum expectation value (VEV) of the radion field, $\phi_{0}$ is the canonically normalized massless radion field, $\hat{H}$ is the Higgs field in the 5D context before rescaling to canonical normalization on the brane. With $\xi \ne 0$, there is neither a pure Higgs boson nor pure radion mass eigenstate. The $T^{\mu\nu}$ is the energy-momentum tensor, which is given at the tree level \cite{csa}
\begin{equation}
T_{\mu \nu}=T_{\mu \nu}^{S M}+T_{\mu \nu}^{D M},
\end{equation}
where
\begin{equation}
\begin{aligned}
T_{\mu \nu}^{S M}= & {\left[\frac{i}{4} \bar{\psi}\left(\gamma_\mu D_\nu+\gamma_\nu D_\mu\right) \psi-\frac{i}{4}\left(\gamma_\mu D_\nu \bar{\psi} \gamma_\mu+D_\mu \bar{\psi} \gamma_\nu\right) \psi-\eta_{\mu \nu}\left(\bar{\psi} \gamma^\mu D_\mu \psi-m_\psi \bar{\psi} \psi\right)+\right.} \\
& \left.+\frac{i}{2} \eta_{\mu \nu} \partial^\rho \bar{\psi} \gamma_\rho \psi\right]+\left[\frac{1}{4} \eta_{\mu \nu} F^{\lambda \rho} F_{\lambda \rho}-F_{\mu \lambda} F_\nu^\lambda\right]+\left[\eta_{\mu \nu} D^\rho H^{\dagger} D_\rho H+\eta_{\mu \nu} V(H)+\right. \\
& \left.+D_\mu H^{\dagger} D_\nu H+D_\nu H^{\dagger} D_\mu H\right],
\end{aligned}
\end{equation}
and
\begin{equation}
T_{\mu \nu}^{D M}=\left(\partial_\mu S\right)\left(\partial_\nu S\right)-\frac{1}{2} \eta_{\mu \nu}\left(\partial^\rho S\right)\left(\partial_\rho S\right)+\frac{1}{2} \eta_{\mu \nu} m_S^2 S^2,
\end{equation}
\begin{equation}
T^{\mu}_{\mu}=\Sigma_{f} m_{f} \overline{f}f - 2m^{2}_{W}W^{+}_{\mu}W^{-\mu}-m^{2}_{Z}Z_{\mu}Z^{\mu} + (2m^{2}_{h_{0}}h_{0}^{2} -  \partial_{\mu}h_{0}\partial^{\mu}h_{0}) + ...
\end{equation}
The states that diagonalize the kinetic energy and have canonical normalization $h$ and $\phi$ are given by \cite{dominici}
\begin{equation} 
\left(\begin{array}{c} {h_{0} } \\ {\phi _{0} } \end{array}\right)=\left(\begin{array}
{cc} {1} & {6\xi \gamma /Z} \\ {0} & {-1/Z} \end{array}\right)\left(\begin{array}{cc}
 {\cos \theta } & {\sin \theta } \\ {-\sin \theta } & {\cos \theta } \end{array}\right)
 \left(\begin{array}{c} {h} \\ {\phi } \end{array}\right)=\left(\begin{array}{cc}
  {d} & {c} \\ {b} & {a} \end{array}\right)\left(\begin{array}{c} {h} \\ {\phi } \end{array}\right), \label{pt1}
\end{equation}
where
$Z^{2} = 1 + 6\gamma ^{2} \xi \left(1 -\, \, 6\xi \right) = \beta - 36\xi ^{2}\gamma ^{2}$ is the coefficient of the radion kinetic term after undoing the kinetic mixing, $\gamma = \upsilon /\Lambda _{\phi }, \upsilon = 246$ GeV. The parameters $a, b, c,d$ define the mixing between the $\xi = 0$ states and the $\xi \neq 0$ mass eigenstates \cite{boo}. \\
\hspace*{1cm}The mixing angle $\theta $ is
\begin{equation}
\tan 2{\theta } = 12{\gamma \xi Z}\frac{m_{h_{0}}^{2}}{m_{h_{0}}^{2} \left( Z^{2} - 36\xi^{2} \gamma ^{2} \right) - m_{\phi _{0}}^{2}},
\end{equation}
where $m_{h_{0}}$ and $m_{\phi _{0}}$ are the Higgs and radion masses before mixing.\\
\hspace*{1cm}The new physical fields h and $\phi $ in (\ref{pt1}) are Higgs-dominated state and radion, respectively \cite{bha}:
\begin{equation} 
m_{h,\phi }^{2} =\frac{1}{2Z^{2} } \left[m_{\phi _{0} }^{2} +\beta m_{h_{0} }^{2} \pm \sqrt{(m_{\phi _{0} }^{2} +\beta m_{h_{0} }^{2} )^{2} -4Z^{2} m_{\phi _{0} }^{2} m_{h_{0} }^{2} } \right].
\end{equation}
\hspace*{1cm}There are four independent parameters $\Lambda _{\phi } ,\, \, m_{h} ,\, \, m_{\phi } ,\, \, \xi$ that must be specified to fix the state mixing parameters. We consider the case of $\Lambda _{\phi } = 5$ TeV and $\frac{m_{0} }{M_{P} } = 0.1$, which makes the radion stabilization model most natural \cite{dav}. It is the worth noting that the $\gamma \gamma$ final state is of particular importance for constraining the model when $\xi$ is near the conformal limit of $\xi = 1/6$ \cite{ahm}.\\ 
\subsection{The scalar unparticle}
\hspace*{1cm}
 The effects of unparticle on properties of high energy colliders have been intensively studied in Refs. \cite{dong, pra, alan, maj, kuma, sahi, kiku, chen, kha, fried}. In the rest of this work, we restrict ourselves by considering only scalar unparticle. The scalar unparticle propagator is given by \cite{georgi2, georgi}
\begin{equation}
\Delta_{scalar} = \dfrac{iA_{d_{U}}}{2sin(d_{U}\pi)}(-q^{2})^{d_{U}-2},
\end{equation}
where 
\begin{align}
&A_{d_{U}} = \dfrac{16\pi^{2}\sqrt{\pi}}{(2\pi)^{2d_{U}}}\dfrac{\Gamma\left( d_{U} + \dfrac{1}{2}\right) }{\Gamma(d_{U} - 1)\Gamma (2d_{U})},\\
&(-q^{2})^{d_{U}-2} = \begin{cases}
|q^{2}|^{d_{U} - 2} e^{-d_{U}\pi}   \text{   for s-channel process}, q^{2}  \text{ is positive,}\\
|q^{2}|^{d_{U} - 2}                \text{   for u-, t-channel process}, q^{2}  \text{ is negative.}
\end{cases}
\end{align}
\hspace*{1cm}The effective interactions for the scalar unparticle operators to fermion and Higgs boson are given by
\begin{equation}
\lambda_{ff}\dfrac{1}{\Lambda^{d_{U}-1}_{U}}\overline{f}f O_{U}, \lambda_{hh}\dfrac{1}{\Lambda^{d_{U}-2}_{U}}H^{+}H O_{U}.
\end{equation}
Feynman rules for the couplings of the scalar unparticle in the RS model are showed as follows \cite{giang}
\begin{align}
&g_{f\overline{f}U} = i\overline{g}_{f\overline{f}U} = i\dfrac{\lambda_{ff}}{\Lambda_{U}^{d_{U} - 1}},\\
&g_{Uhh} = -i\overline{g}_{Uhh} = -i\dfrac{\lambda_{hh}}{\Lambda_{U}^{d_{U} - 2}},\\
&g_{U\phi\phi} = - i\overline{g}_{U\phi\phi} = -i\dfrac{\lambda_{\phi\phi}}{\Lambda_{U}^{d_{U} - 2}}.  
\end{align}
\hspace*{1cm}Using the above formulas and Feynman vertex for the couplings of Higgs/radion given in detail in Appendix A, we will study the influence of  the scalar unparticle and anomalous couplings at muon colliders in final states with  multiple photons.
\section{The influence of the scalar unparticle and anomalous couplings on the production of the multiple photons}
\hspace*{1cm} Bounds from LEP on unparticle interactions with electroweak bosons in which four photon signals are investigated in Ref.\cite{scott}. The multi - photon signals at LHC through the unparticle self-interaction have been studied in detail in Ref.\cite{alie}. Searchs for new physics in final states with multiple leptons in high energy collisions are considered in detail\cite{alb}. Recently, the search for the exotic decay of the Higgs boson into two light pseudoscalars with four photons in  final states at the high energy colliders has been presented\cite{arm}. The new physical phenomenology concerning the lepton colliders is presented  in detail in Refs.\cite{arnab, fridell, antonov, Li, Lu}. In this work, we will evaluate  the  influence of the scalar unparticle and anomalous couplings at muon colliders, which shows that cross-section for the production of multiple photons in the final states provides an important signature  for new phenomena  in the Randall-Sundrum model. To be consistent with current experiments, the collision energies in our calculations are fixed in range of $7$ TeV to $14$ TeV as in Refs.\cite{alie,cms16,arm}.\\
\subsection{The $\mu^{+}\mu^{-} \rightarrow hh/\phi\phi \rightarrow\gamma\gamma\gamma\gamma$ collisions}
\begin{figure}[!htb] 
\begin{center}
\begin{tabular}{cc}
        \includegraphics[width=8cm, height= 5cm]{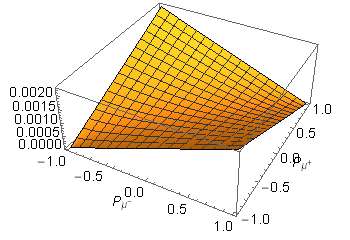} &
        \includegraphics[width=7.5cm, height= 5cm]{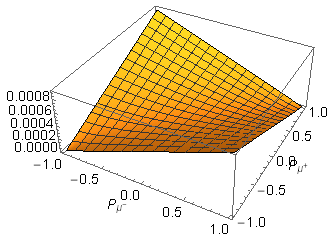} \\
        (a) & (b)
    \end{tabular}        
             \caption{\label{Fig.2} The total cross-section as a function of the polarization coefficients of muon and antimuon beam in (a) $\mu^{+}\mu^{-} \rightarrow hh \rightarrow \gamma\gamma\gamma\gamma$, (b) $\mu^{+}\mu^{-} \rightarrow \phi\phi \rightarrow \gamma\gamma\gamma\gamma$ collisions. The parameters are chosen as $\sqrt{s} = 8$ TeV, $d_{U} = 1.1$, $m_{\phi} = 110$ GeV, $\Lambda_{\phi} = 5$ TeV, $\Lambda_{U} = 1$ TeV. }
\end{center}
\end{figure}

\begin{figure}[!htb] 
\begin{center}
           \begin{tabular}{cc}
        \includegraphics[width=7.5cm, height= 5cm]{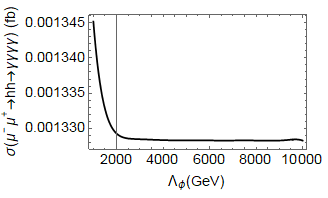} &
        \includegraphics[width=8cm, height= 5cm]{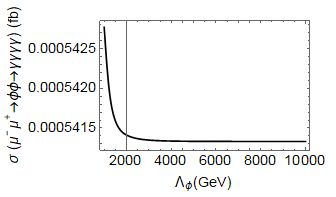} \\
        (a) & (b)
    \end{tabular}    
    \caption{\label{Fig.3} The total cross-section as a function of the $\Lambda_{\phi}$ in (a) $\mu^{+}\mu^{-} \rightarrow hh \rightarrow \gamma\gamma\gamma\gamma$, (b) $\mu^{+}\mu^{-} \rightarrow \phi\phi \rightarrow \gamma\gamma\gamma\gamma$ collisions. The parameters are taken to be $P_{\mu^{-}} = 0.8, P_{\mu^{+}} = - 0.3$, $\sqrt{s} = 8$, $d_{U} = 1.1$, $m_{\phi} = 110$ GeV, $\Lambda_{U} = 1$ TeV. }
        \end{center}
\end{figure}

\begin{figure}[!htb] 
\begin{center}
\begin{tabular}{cc}
        \includegraphics[width=7.5cm, height= 5cm]{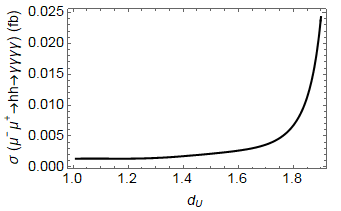} &
        \includegraphics[width=8cm, height= 5cm]{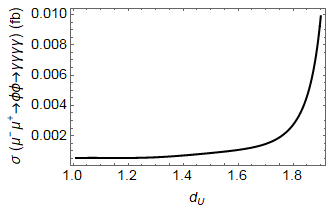} \\
        (a) & (b)
    \end{tabular}
    \caption{\label{Fig.4} The total cross-section as a function of $d_{U}$ in (a) $\mu^{+}\mu^{-} \rightarrow hh \rightarrow \gamma\gamma\gamma\gamma$, (b) $\mu^{+}\mu^{-} \rightarrow \phi\phi \rightarrow \gamma\gamma\gamma\gamma$ collisions. The parameters are taken to be $P_{\mu^{-}} = 0.8, P_{\mu^{+}} = - 0.3$, $\sqrt{s} = 8$ TeV, $m_{\phi} = 110$ GeV, $\Lambda_{\phi} = 5$ TeV, $\Lambda_{U} = 1$ TeV. }
        \end{center}
\end{figure}

\begin{figure}[!htb] 
\begin{center}
\begin{tabular}{cc}
        \includegraphics[width=7.5cm, height= 5cm]{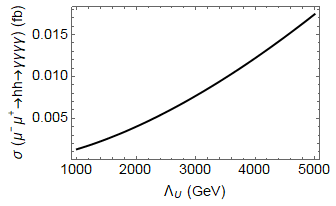} &
        \includegraphics[width=8cm, height= 5cm]{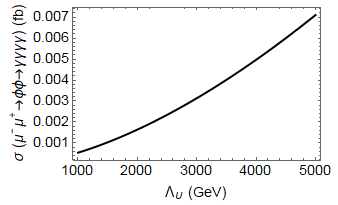} \\
        (a) & (b)
    \end{tabular}
         \caption{\label{Fig.5} The total cross-section as a function of $\Lambda_{U}$ in (a) $\mu^{+}\mu^{-} \rightarrow hh \rightarrow \gamma\gamma\gamma\gamma$, (b) $\mu^{+}\mu^{-} \rightarrow \phi\phi \rightarrow \gamma\gamma\gamma\gamma$ collisions. The parameters are taken to be $P_{\mu^{-}} = 0.8, P_{\mu^{+}} = - 0.3$, $\sqrt{s} = 8$ TeV, $m_{\phi} = 110$ GeV, $\Lambda_{\phi} = 5$ TeV. }
        \end{center}
\end{figure}

\hspace*{1cm} Now, we investigate the phenomenology of a pair of Higgs boson or radion at muon collider, followed by scalar particle decaying into a pair of two  photons. The collision process is considered as
\begin{equation}
\mu^{-}(p_{1}) + \mu^{+}(p_{2}) \    \rightarrow         X (k_{1}) + X (k_{2}).
\end{equation}
Here, $p_{i}, k_{i}$ (i = 1,2) stand for the momentums, respectively. X stands for Higgs or radion. Feymann diagram is given in Figure.\ref{Fig.13} (Appendix B).
The transition amplitude representing s-channel is given by
\begin{equation}
M_{s} = -i\left( \dfrac{\overline{g}_{\mu\mu\phi}\overline{g}_{\phi XX}}{q_{s}^{2} - m^{2}_{\phi}} + \dfrac{\overline{g}_{\mu\mu h}\overline{g}_{hXX}}{q_{s}^{2} - m^{2}_{h}} + \overline{g}_{\mu\mu U}\overline{g}_{UXX} \dfrac{A_{d_{U}}}{2sin(d_{U}\pi)} (-q_{s}^{2})^{d_{U} - 2}\right)\overline{v}(p_{2})u(p_{1}) .
\end{equation}
\hspace*{1cm}The transition amplitude representing u-channel is 
\begin{equation}
M_{u} = -i\overline{g}_{\mu\mu X}\overline{g}_{\mu\mu X}\overline{v}(p_{2})\dfrac{(\slashed{q}_{u} + m_{\mu})}{q_{u}^{2} - m^{2}_{\mu}}u(p_{1}).
\end{equation}
\hspace*{1cm}The transition amplitude representing t-channel is
\begin{equation}
M_{t} = -i\overline{g}_{\mu\mu X}\overline{g}_{\mu\mu X}\overline{v}(p_{2})\dfrac{(\slashed{q}_{t} + m_{\mu})}{q_{t}^{2} - m^{2}_{\mu}}u(p_{1}).
\end{equation}
Here, $ \overline{g}_{\phi hh}, \overline{g}_{hhh}$, $\overline{g}_{\mu\mu\phi}, \overline{g}_{\mu\mu h}$, $\overline{g}_{h\phi\phi}, \overline{g}_{\phi \phi\phi}$  are given by \cite{dominici}. From the formula 19, we can see that there is the contribution of the scalar unparticle propagator in s-channel, which is important in our calculations. The total cross-sections for the production of four photons in final states are calculated as follows
\begin{align}
&\sigma_{hh} = \sigma (\mu^{-} \mu^{+} \rightarrow hh ) \times  2 Br(h\rightarrow \gamma\gamma),\\
&\sigma_{\phi \phi} = \sigma (\mu^{-} \mu^{+} \rightarrow \phi\phi ) \times  2 Br(\phi\rightarrow \gamma\gamma).
\end{align}
where 
\begin{equation}
\frac{d\sigma (\mu^{-} \mu^{+} \rightarrow  XX)}{dcos\psi} = \frac{1}{32 \pi s} \frac{|\overrightarrow{k}_{1}|}{|\overrightarrow{p}_{1}|} |M_{fi}|^{2}
\end{equation}
is the expressions of the differential cross-section \cite{pes}. Based on the above formulas, with the model parameters are chosen as $\Lambda_{\phi} = 5$ TeV, $ m_{h} = 125$ GeV, $m_{\phi} = 110$ GeV \cite{soael}, $\lambda_{\mu\mu} = \lambda_{hh} = \lambda_{\phi\phi} = \lambda_{0} = 1$, $\Lambda_{U} = 1$ TeV, $1 < d_{U} < 2$ (in case of the scalar unparticle \cite{fried}), we give estimates in detail  for the cross-sections as follows \\
\hspace*{1cm} i) The total cross-sections depend on polarization coefficients  shown in Fig.\ref{Fig.2}. Here, $P_{\mu^{-}}, P_{\mu^{+}}$ are the polarization coefficients of the muon and antimuon beams, respectively. The scaling dimension of the unparticle operator has been taken to be $d_{U} = 1.1$. From the figure we can see that the total cross-sections achieve the minimum value in case of $P_{\mu^{-}} = P_{\mu^{+}} = \pm 1$ and the maximum value in case of $P_{\mu^{-}} = -1, P_{\mu^{+}} = 1$ and vice versa.\\
\hspace*{1cm} ii) The total cross-sections depend on the VEV of the radion field are shown in Fig.\ref{Fig.3} which shows that cross-sections decrease fast in the region 1 TeV $\leq \Lambda_{\phi} \leq $ 2 TeV. From the figure we can see that the cross sections are flat when  $\Lambda_{\phi} > 2$ TeV.\\
\hspace*{1cm} iii) In Fig.\ref{Fig.4}, we plot the total cross-sections as the function of $d_{U}$ in the cases $P_{\mu^{-}} = 0.8, P_{\mu^{+}} = -0.3$\cite{bsung1, Clic1}. From the figure we can see that the cross-sections increase as $d_{U}$ increases.\\
\hspace*{1cm} iv) In Fig.\ref{Fig.5}, we plot the total cross-sections as the function of $\Lambda_{U}$. From the figure we can see that the cross-sections increase as $\Lambda_{U}$ increases. With $d_{U} = 1.1$, $\Lambda_{U} = 1$ TeV, the total cross-section for production of four photons is consistent with the result in Ref. \cite{alie}.  

\subsection{The $\mu^{+} \mu^{-} \rightarrow Uh/ U\phi \rightarrow U \gamma \gamma$ collisions}
\begin{figure}[!htb] 
\begin{center}   
\begin{tabular}{cc}
        \includegraphics[width=8cm, height= 5cm]{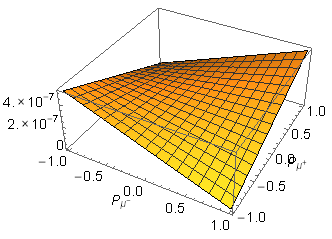} &
        \includegraphics[width=8cm, height= 5cm]{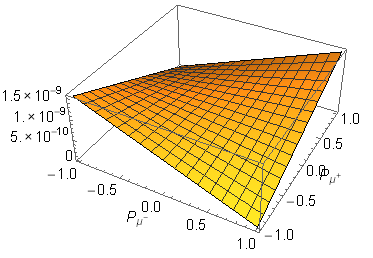} \\
        (a) & (b)
    \end{tabular}        
       \caption{\label{Fig.6} The total cross-section as a function of the polarization coefficients of muon and antimuon beam in (a) $\mu^{+}\mu^{-} \rightarrow Uh \rightarrow U\gamma\gamma$, (b) $\mu^{+}\mu^{-} \rightarrow U\phi \rightarrow U\gamma\gamma$ collisions. The parameters are chosen as $\sqrt{s} = 8$ TeV, $d_{U} = 1.1$, $\Lambda_{\phi} = 5$ TeV, $\Lambda_{U} = 1$ TeV. }
        \end{center}
\end{figure}

\begin{figure}[!htb] 
\begin{center}    
\begin{tabular}{cc}
        \includegraphics[width=7cm, height= 5cm]{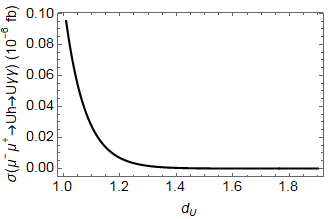} &
        \includegraphics[width=8cm, height= 5cm]{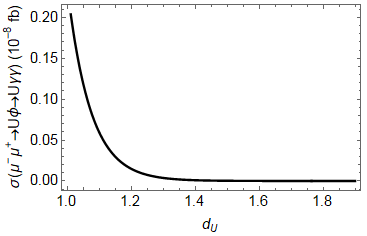} \\
        (a) & (b)
    \end{tabular}         
              \caption{\label{Fig.8} The total cross-section as a function of $d_{U}$ in (a) $\mu^{+}\mu^{-} \rightarrow Uh \rightarrow U\gamma\gamma$, (b) $\mu^{+}\mu^{-} \rightarrow U\phi \rightarrow U\gamma\gamma$ collisions. The parameters are taken to be $P_{\mu^{-}} = 0.8, P_{\mu^{+}} = - 0.3$, $\sqrt{s} = 8$ TeV, $m_{\phi} = 110$ GeV, $\Lambda_{\phi} = 5$ TeV, $\Lambda_{U} = 1$ TeV. }
        \end{center}
\end{figure}

\begin{figure}[!htb] 
\begin{center}    
\begin{tabular}{cc}
      \includegraphics[width=7cm, height= 5cm]{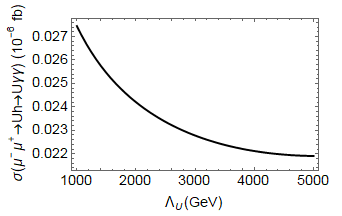} &
       \includegraphics[width=8cm, height= 5cm]{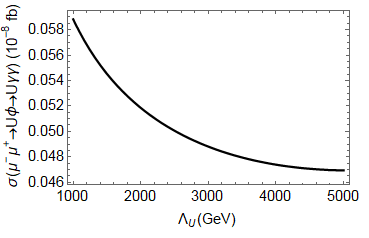} \\
       (a) & (b)
    \end{tabular}         
             \caption{\label{Fig.9} The total cross-section as a function of the energy scale $\Lambda_{U}$ in (a) $\mu^{+}\mu^{-} \rightarrow Uh \rightarrow U\gamma\gamma$, (b) $\mu^{+}\mu^{-} \rightarrow U\phi \rightarrow U\gamma\gamma$ collisions. The parameters are taken to be $P_{\mu^{-}} = 0.8, P_{\mu^{+}} = - 0.3$, $d_{U} = 1.1$, $\sqrt{s} = 8$ TeV, $m_{\phi} = 110$ GeV. }
        \end{center}
   \end{figure}

\hspace*{1cm}Next, we study the $\mu^{-} \mu^{+} \rightarrow Uh/ U\phi \rightarrow U \gamma \gamma$ processes in which the initial state contains muon and antimuon, the final state contains a pair of unparticle and the scalar particle (Higgs/radion) which decay into two photons.
\begin{equation}
\mu^{-}(p_{1}) + \mu^{+}(p_{2}) \    \rightarrow         U (k_{1}) + X (k_{2}),
\end{equation}
Here, X stands for Higgs or radion, $p_{i}, k_{i}$ (i = 1,2) stand for the momentums, respectively. Feynman diagram is shown in Figure.\ref{Fig.14} (Appendix B). The transition amplitude representing s - channel is given by 
\begin{equation}
M_{s} = i \dfrac{\overline{g}_{\mu\mu X}\overline{g}_{UXX}}{q_{s}^{2} - m^{2}_{X}}\overline{v}(p_{2})u(p_{1}).
\end{equation}
\hspace*{1cm}The transition amplitude representing u - channel is 
\begin{equation}
M_{u} = - i\dfrac{\overline{g}_{\mu\mu X}\overline{g}_{\mu\mu U}}{q_{u}^{2} - m^{2}_{\mu}}\overline{v}(p_{2})(\slashed{q}_{u} + m_{\mu}) u(p_{1}).
\end{equation}
\hspace*{1cm}The transition amplitude representing t-channel is 
\begin{equation}
M_{t} = - i\dfrac{\overline{g}_{\mu\mu X}\overline{g}_{\mu\mu U}}{q_{t}^{2} - m^{2}_{\mu}}u(p_{1})(\slashed{q}_{t} + m_{\mu})\overline{v}(p_{2}).
\end{equation}
The total cross-sections are given by
\begin{align}
&\sigma_{Uh} = \sigma (\mu^{-} \mu^{+} \rightarrow Uh ) \times   Br(h\rightarrow \gamma\gamma ),\\
&\sigma_{U\phi} = \sigma (\mu^{-} \mu^{+} \rightarrow U\phi ) \times  Br(\phi \rightarrow \gamma\gamma).
\end{align}
\hspace*{1cm} 
 Now we give some estimates in detail for the cross-sections as follows \\
\hspace*{1cm} i) With the parameters are chosen as in Fig.1, the total cross-sections depend on  polarization coefficients shown in Fig.\ref{Fig.6}. From the figure we can see that the cross-sections achieve the maximum value in case of $P_{\mu^{-}} =  P_{\mu^{+}} = \pm 1$ and the minimum value in case of $P_{\mu^{-}} =  1$, $P_{\mu^{+}} = - 1$ and vice versa.\\
\hspace*{1cm} ii) In Fig.\ref{Fig.8} we plot the total cross-sections as the function of $d_{U}$ in case of fixed polarization coefficients, $P_{\mu^{-}} = 0.8, P_{\mu^{+}} = -0.3$. From the figure we can see that the cross sections decrease rapidly as $d_{U}$ increases and they are flat when  $d_{U} > 1.4 $. \\
\hspace*{1cm} iii) Cross-sections depend on the $\Lambda_{U}$ shown in Fig.\ref{Fig.9} which shows that cross-sections decrease as $\Lambda_{U}$ increases. It is the worth noting that the behavior of the cross-section in the figures 6, 7 is contrary to the results obtained from the figures 3, 4 in previous process, respectively.\\
 \subsection{The $\mu^{+}\mu^{-} \rightarrow Zh/Z\phi\rightarrow \gamma\gamma\gamma\gamma$ collisions}
\begin{figure}[!htb] 
\begin{center}   
\begin{tabular}{cc}
        \includegraphics[width=8cm, height= 5cm]{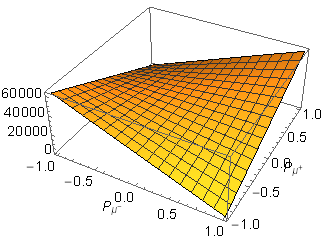} &
        \includegraphics[width=8cm, height= 5cm]{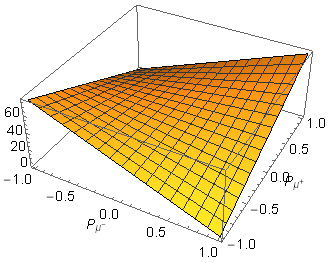} \\
        (a) & (b)
    \end{tabular}        
       \caption{\label{Fig.10} The total cross-section as a function of the polarization coefficients of muon and antimuon beam in (a) $\mu^{+}\mu^{-} \rightarrow Zh \rightarrow \gamma\gamma\gamma\gamma$, (b) $\mu^{+}\mu^{-} \rightarrow Z\phi \rightarrow \gamma\gamma\gamma\gamma$ collisions. The parameters are taken to be $\sqrt{s} = 8$ TeV, $m_{\phi} = 110$ GeV, $\Lambda_{\phi} = 5$ TeV. }
        \end{center}
\end{figure}

\begin{figure}[!htb] 
\begin{center}    
\begin{tabular}{cc}
        \includegraphics[width=8cm, height= 5cm]{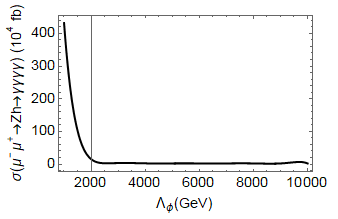} &
        \includegraphics[width=8cm, height= 5cm]{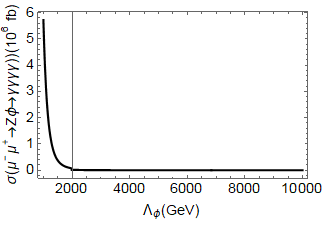} \\
        (a) & (b)
    \end{tabular}         
              \caption{\label{Fig.11} The total cross-section as a function of $\Lambda_{\phi}$ in (a) $\mu^{+}\mu^{-} \rightarrow Zh \rightarrow \gamma\gamma\gamma\gamma$, (b) $\mu^{+}\mu^{-} \rightarrow Z\phi \rightarrow \gamma\gamma\gamma\gamma$ collisions. The parameters are taken to be $P_{\mu^{-}} = 0.8, P_{\mu^{+}} = - 0.3$, $\sqrt{s} = 8$ TeV, $m_{\phi} = 110$ GeV. }
        \end{center}
\end{figure}

\begin{figure}[!htb] 
\begin{center}    
\begin{tabular}{cc}
        \includegraphics[width=8cm, height= 5cm]{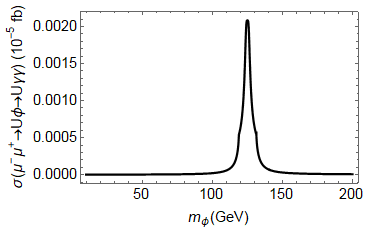} &
        \includegraphics[width=8cm, height= 5cm]{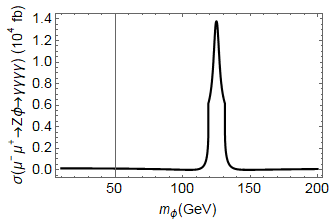} \\
        (a) & (b)
    \end{tabular}         
              \caption{\label{Fig.12} The total cross-section as a function of $m_{\phi}$ in (a) $\mu^{+}\mu^{-} \rightarrow U\phi \rightarrow U\gamma\gamma$, (b) $\mu^{+}\mu^{-} \rightarrow Z\phi \rightarrow \gamma\gamma\gamma\gamma$ collisions. The parameters are taken to be $P_{\mu^{-}} = 0.8, P_{\mu^{+}} = - 0.3$, $\sqrt{s} = 8$ TeV, $\Lambda_{\phi} = 5$ TeV. }
        \end{center}
\end{figure}

\begin{figure}[!htbp] 
\begin{center}
\includegraphics[width= 14 cm,height= 4 cm]{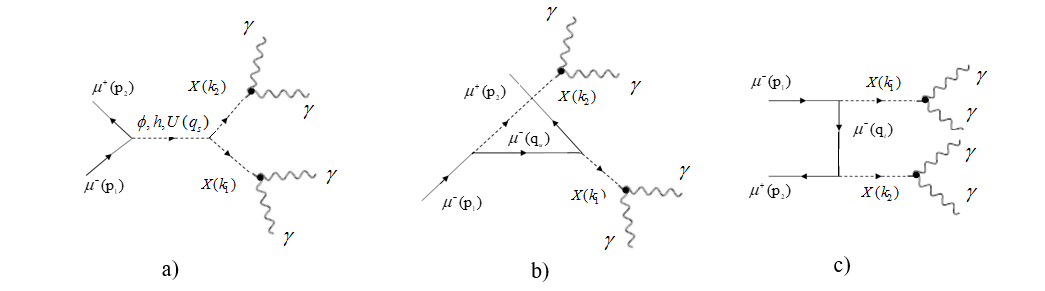}
\caption{\label{Fig.13} Feynman diagrams for $\mu^{+}\mu^{-} \rightarrow XX \rightarrow \gamma\gamma\gamma\gamma$ collisions. X stands for the Higgs or radion. The figures (a), (b), (c) representing the s, u, t-channel exchange, respectively.}
\end{center}
\end{figure}

\begin{figure}[!htbp] 
\begin{center}
\includegraphics[width= 14 cm,height= 4 cm]{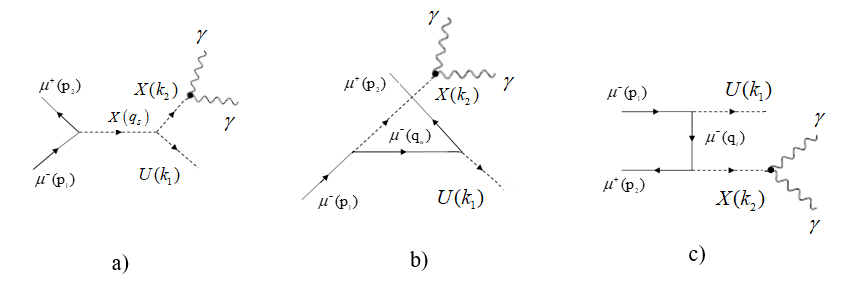}
\caption{\label{Fig.14} Feynman diagrams for $\mu^{+}\mu^{-} \rightarrow UX \rightarrow U\gamma\gamma$ collisions. X stands for the Higgs or radion. The figures (a), (b), (c) representing the s, u, t-channel exchange, respectively.}
\end{center}
\end{figure}

\begin{figure}[!htbp] 
\begin{center}
\includegraphics[width= 14 cm,height= 4 cm]{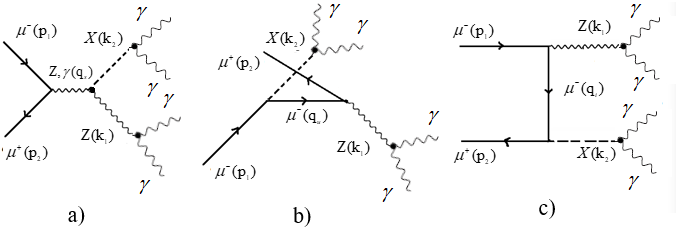}
\caption{\label{Fig.15} Feynman diagrams for $\mu^{+}\mu^{-} \rightarrow ZX \rightarrow \gamma\gamma\gamma\gamma$ collisions. X stands for the Higgs or radion. The figures (a), (b), (c) representing the s, u, t-channel exchange, respectively.}
\end{center}
\end{figure}

\hspace*{1cm}Finally, we consider the collision process in which the initial state contains muon and antimuon, the final state contains the four photons through scalar anomalous couplings
 \begin{equation}
\mu^{-}(p_{1}) + \mu^{+}(p_{2}) \rightarrow Z (k_{1}) + X (k_{2}).
\end{equation}
Here, X stands for Higgs or radion. In the final state, $Zh/Z\phi$ decay into four photons. Feynman diagram with the contribution of scalar anomalous couplings is given in Figure.\ref{Fig.15} (Appendix B). The transition amplitude representing s-channel can be written as follows
\begin{equation}
 M_{s} = M_{Z} + M_{\gamma},
\end{equation}
where
\begin{equation}
M_{Z} = \frac{{ {{\bar g}_{ZZX}}}}{{q_s^2 - m_Z^2}}\bar v({p_2}){\gamma ^\mu }\left( {{v_\mu} - {a_\mu}{\gamma ^5}} \right)u({p_1})\left( {{\eta _{\mu \beta }} - \frac{{{q_{s\mu }}{q_{s\beta }}}}{{m_Z^2}}} \right)\left[ {{\eta ^{\beta \nu }} - 2g_\phi ^Z\left( {{\eta ^{\beta \nu }}{k_1}{q_s} - q_s^\nu k_1^\beta } \right)} \right]\varepsilon _\nu ^*({k_1}),
\end{equation}
\begin{equation}
M_{\gamma } = \frac{{ - {C_{\gamma ZX }}}}{{q_s^2}}\bar v({p_2}){\gamma ^\mu }\left( {{v_\mu} - {a_\mu}{\gamma ^5}} \right)u({p_1}){\eta _{\mu \beta }}\left[ {{\eta ^{\beta \nu }}{k_1}{q_s} - q_s^\nu k_1^\beta } \right]\varepsilon _\nu ^*({k_1}).
\end{equation}
\hspace*{1cm}The transition amplitude representing u-channel is given by
\begin{equation}
M_{u} = \frac{-i\overline{g}_{\mu Z}\overline{g}_{\mu\mu X}}{(q_{u}^{2} - m_{\mu}^{2})} \varepsilon^{*}_{\mu} (k_{1}) \overline{v}(p_{2}) \gamma^{\mu} (v_{\mu} - a_{\mu}\gamma^{5}) (\widehat{q}_{u} + m_{\mu})u(p_{1}).
\end{equation}
\hspace*{1cm}The transition amplitude representing t-channel is given by
\begin{equation}
M_{t} = \frac{-i\overline{g}_{\mu Z}\overline{g}_{\mu\mu X}}{(q_{t}^{2} - m_{\mu}^{2})} \varepsilon^{*}_{\mu} (k_{1}) \overline{v}(p_{2}) \gamma^{\mu} (v_{\mu} - a_{\mu}\gamma^{5}) (\widehat{q}_{t} + m_{\mu})u(p_{1}).
\end{equation}
Here, $\overline{g}_{ZZh}, \overline{g}_{ZZ\phi}, C_{\gamma Zh}, C_{\gamma Z\phi}$ are given by \cite{ahm}. The total cross-section for the above processes is written as follows 
\begin{equation}
\sigma_{ZX} = \sigma (\mu^{-} \mu^{+} \rightarrow ZX ) \times   Br(Z\rightarrow \gamma\gamma) \times   Br(X\rightarrow \gamma\gamma ).
\end{equation}
\hspace*{1cm} Using the model parameters as in the previous sections and branching ratio of Z boson in Ref. \cite{work}, we have estimates in detail for the cross-sections as follows \\
\hspace*{1cm} i) The total cross-sections depend on typical polarization coefficients is shown in Fig.\ref{Fig.10}. The total cross-sections achieve the maximum value in case of $P_{\mu^{-}} =  P_{\mu^{+}} = \pm 1$ and the minimum value in case of $P_{\mu^{-}} =  1$, $P_{\mu^{+}} = - 1$ and vice versa. This result is similar to the previous processes. \\
 \hspace*{1cm} ii) The total cross-sections depend on the $\Lambda_{\phi}$ are shown in Fig.\ref{Fig.11}. From the figure  we can see that the total cross-sections decrease fast in the region 1 TeV $\leq \Lambda_{\phi} \leq $ 2 TeV, gradually in the region 2 TeV $\leq \Lambda_{\phi} \leq $ 10 TeV.\\
 \hspace*{1cm} iii) In Fig.\ref{Fig.12}, we plot the total cross-sections as the function of the radion mass $m_{\phi}$ in the cases $\mu^{+}\mu^{-} \rightarrow U\phi \rightarrow U\gamma\gamma$ and $\mu^{+}\mu^{-} \rightarrow Z\phi \rightarrow \gamma\gamma\gamma\gamma$ collisions, respectively. From the figure, we can see that the cross sections achieve the maximum value at the radion-dominated state, $m_{\phi} = 125$ GeV.\\
\hspace*{1cm} iv) Some typical values for the total cross-sections with the different collision energy values are given in detail in Table \ref{tab1} and Table \ref{tab2}.\\
 \hspace*{1cm}  The numerical values from Table \ref{tab1} show that with the contribution of scalar unparticle, the cross-sections for the production of four-photons are much larger than the associated production of two-photons with the scalar unparticle. This is due to the scalar unparticle propagator in the s-channel gives main contribution on the considered processes. The numerical values in the processes without the unparticle interaction are shown in Table \ref{tab2}. From results, we can see that the cross-sections for the production of four photons with the contribution of scalar anomalous couplings are much larger than that of unparticle under the same conditions. It is the worth noting that the branching ratio of Higgs boson is larger than that of radion \cite{vic, gicho}, so the total cross-sections through $hh$, $Uh, Zh$ production are much larger than that through $\phi\phi$, $U\phi, Z\phi$, respectively. In the four-photons final state, the cross-sections through $Zh/Z\phi$ are much larger than that through $hh/\phi\phi$, due to the strength of the scalar anomalous couplings $\gamma Zh/\gamma Z\phi$.\\
 
 	  \begin{table}[!htb]
 	  	  \caption{\label{tab1} Some typical values for production cross-sections of multi - photons with the different values $\sqrt{s}$. The parameters are chosen as $P_{\mu^{-}} = 0.8 $, $P_{\mu^{+}} = -0.3$, $\xi = 1/6$, $\Lambda_{\phi} = 5$ TeV, $m_{\phi} = 125$ GeV, $\Lambda_{U} = 1$ TeV, $\lambda_{0} = 1$.} 
 	  	  \vspace*{0.5cm}
 	  	\begin{center}
 	   	  \begin{tabular}{|c|c|c|c|c|}   
	\hline 
	$\sqrt{s}$ (TeV) & 7 & 8 & 13 & 14  \\ 
	\hline
	$\sigma_{\mu^{+}\mu^{-} \rightarrow U\phi \rightarrow U\gamma\gamma} (10^{-5} fb)$
	&0.00272 & 0.00208 & 0.00078 & 0.00068 \\
	\hline
	$\sigma_{\mu^{+}\mu^{-} \rightarrow Uh \rightarrow U\gamma\gamma} (10^{-5} fb)$
	& 0.00831 & 0.00635 & 0.00240 & 0.00207  \\
	\hline 
	$\sigma_{\mu^{+}\mu^{-} \rightarrow \phi\phi \rightarrow \gamma\gamma\gamma\gamma} (fb)$
	& 0.00087 & 0.00054 & 0.00009 & 0.00007  \\
	\hline
	$\sigma_{\mu^{+}\mu^{-} \rightarrow hh \rightarrow \gamma\gamma\gamma\gamma} (fb)$
	& 0.00214 & 0.00132 & 0.00023 & 0.00018\\ 
	\hline
    \end{tabular}
    \end{center}
   \end{table}
   
	 \begin{table}[!htb]
 	  	  \caption{\label{tab2} Some typical values for production cross-sections with the different values $\sqrt{s}$ without the unparticle interaction. The parameters $P_{\mu^{-}}$, $P_{\mu^{+}}$, $\xi$, $\Lambda_{\phi}$, $m_{\phi}$ are chosen as in Table 1.} 
 	  	  \vspace*{0.5cm}
 	  	\begin{center}
 	   	  \begin{tabular}{|c|c|c|c|c|} 
\hline 
	$\sqrt{s}$ (TeV) & 7 & 8 & 13 & 14  \\  	   	
 	   	  \hline
 	   	  $\sigma_{\mu^{+}\mu^{-} \rightarrow \phi\phi \rightarrow \gamma\gamma\gamma\gamma} (10^{-6} fb)$
	& 0.015& 0.012 & 0.007 & 0.006 \\
	\hline
	$\sigma_{\mu^{+}\mu^{-} \rightarrow hh \rightarrow \gamma\gamma\gamma\gamma} (10^{-6} fb)$
	& 2.891 & 2.890 & 2.889 & 2.888\\ 
	\hline
	$\sigma_{\mu^{+}\mu^{-} \rightarrow Z\phi \rightarrow \gamma\gamma\gamma\gamma} (10^{2} fb)$
	&137.861 & 137.876 & 137.905 & 137.907 \\
	\hline
	$\sigma_{\mu^{+}\mu^{-} \rightarrow Zh \rightarrow \gamma\gamma\gamma\gamma} (10^{2} fb)$
	& 482.481 & 482.531 & 482.632 & 482.64  \\
	\hline
    \end{tabular}
    \end{center}
   \end{table}

\section{Conclusion}
\hspace*{1cm} In this paper, we have evaluated the influence of the scalar  unparticle and  anomalous couplings at muon colliders in final states with  multiple photons in the Randall- Sundrum model. The results indicate that with fixed collision energies the total cross-sections for the production of multiple photons depend strongly on the polarization of the muon beams, the parameters of unparticle physics (the scaling dimension $d_{U}$,  operator $\mathcal{O}_{U}$, the energy scale $\Lambda_{U}$) and also the strength of  anomalous couplings. Numerical evaluation shows that  with the contribution of scalar unparticle, the cross-sections for the production of four-photons are much larger than the associated production of two-photons with the scalar unparticle. The cross-sections for the production of {\bf four photons} in finale states with the contribution of scalar anomalous couplings are much larger than that of the unparticle under the same conditions. In the Higgs-radion mixing, the cross sections achieve the maximum value at the radion-dominated state, $m_{\phi} = 125$ GeV, in which the cross-section is much enhanced and can be  measurable  in current experiments. With the integrated luminosity $L = 120 fb^{-1}$\cite{Lu}, one expects several thousand events.\\
 \hspace*{1cm}Finally, we note that in this work we have only considered on a theoretical basis, other problems concerning experiments for production of multiple photons, the reader can see in detail in Refs.\cite{alie,cms16,arm}. The other processes concerning anomalous couplings in the RS model will be studied in our future works.\\
 {\bf Acknowledgements}: The work is supported in part by the National Foundation for Science and Technology Development (NAFOSTED) of Vietnam under Grant No. 103.01-2023.50.\\

\newpage
\newpage
\newpage
\newpage

\newpage
\newpage

\begin{center}
 { APPENDIX A: Feynman vertex for the couplings of Higgs/radion and scalar  anomalies  in Randall-Sundrum model}\\
 \end{center}
 Feynman rules for the couplings of Higgs and radion are showed as follows
\begin{align}
&V(h, \mu^{+}, \mu^{-}) = i\overline{g}_{\mu\mu h} = -i\dfrac{gm_{\mu}}{2m_{W}}\left( d + \gamma b\right),\\
&V(\phi, \mu^{+}, \mu^{-}) = i\overline{g}_{\mu\mu\phi} = -i\dfrac{gm_{\mu}}{2m_{W}}\left( c + \gamma a\right),\\
&V(h, h, h) = i\overline{g}_{hhh} = \dfrac{i}{\Lambda_{\phi}}\left[bd\left(\left(12b\gamma \xi + d(6\xi + 1)\right)(k_{1}^{2} + k_{2}^{2} + k_{3}^{2}) - 12d m_{h_{0}}^{2}\right) - 3\gamma^{-1}d^{3}m_{h_{0}}^{2}\right],\\
&V(\phi, \phi, \phi) = i\overline{g}_{\phi\phi\phi} = \dfrac{i}{\Lambda_{\phi}}\left[ac\left(\left(12a\gamma \xi + c(6\xi + 1)\right)(k_{1}^{2} + k_{2}^{2} + k_{3}^{2}) - 12c m_{h_{0}}^{2}\right)- 3\gamma^{-1}c^{3}m_{h_{0}}^{2}\right],
\end{align}
\begin{equation}
\begin{aligned}
V(\phi, \phi, h) = & i\overline{g}_{\phi\phi h}\\
 = & \dfrac{i}{\Lambda_{\phi}}\Biggl[ \left(6a\xi(\gamma(ad + bc) + cd) + bc^{2} \right) (k_{1}^{2} + k_{2}^{2}) +\\
&+ c\left(12ab\gamma \xi + 2ad + bc(6\xi - 1) \right) k_{3}^{2} - 4c(2ad + bc)m_{h_{0}}^{2} - 3\gamma^{-1}c^{2}d m_{h_{0}}^{2}  \Biggr],
\end{aligned}
\end{equation}
\begin{equation}
\begin{aligned}
V(\phi, h, h) = & i\overline{g}_{\phi hh}\\
 = & \dfrac{i}{\Lambda_{\phi}}\Biggl[ \left(6b\xi(\gamma(ad + bc) + cd) + ad^{2} \right) (k_{1}^{2} + k_{2}^{2}) +\\
&+ d\left(12ab\gamma \xi + 2bc + ad(6\xi - 1) \right) k_{3}^{2} - 4d(ad + 2bc)m_{h_{0}}^{2} - 3\gamma^{-1}cd^{2} m_{h_{0}}^{2}  \Biggr],
\end{aligned}
\end{equation}
Feynman vertex for the scalar  anomalous  couplings is given by
\begin{equation}
\begin{aligned}
&V(h, \gamma_{\mu}(k_{1}), \gamma_{\nu}(k_{2})) = iC_{\gamma h}\left[(k_{1}k_{2})\eta^{\mu\nu} - k_{1}^{\nu}k_{2}^{\mu}\right]\\
&= -i\dfrac{\alpha}{2\pi\upsilon_{0}}\left((d + \gamma b)\sum_{i}e_{i}^{2}N_{c}^{i}F_{i}(\tau_{i}) - (b_{2} + b_{Y} + \frac{4\pi}{\alpha kb_{0}})\gamma b \right)\times \left[(k_{1}k_{2})\eta^{\mu\nu} - k_{1}^{\nu}k_{2}^{\mu}\right],
\end{aligned}
\end{equation}

\begin{equation}
\begin{aligned}
&V(\phi, \gamma_{\mu}(k_{1}), \gamma_{\nu}(k_{2})) =
iC_{\gamma \phi}\left[(k_{1}k_{2})\eta^{\mu\nu} - k_{1}^{\nu}k_{2}^{\mu}\right]\\
&=  -i\dfrac{\alpha}{2\pi\upsilon_{0}}\left((c + \gamma a)\sum_{i}e_{i}^{2}N_{c}^{i}F_{i}(\tau_{i}) - (b_{2} + b_{Y} + \frac{4\pi}{\alpha kb_{0}})\gamma a \right) \left[(k_{1}k_{2})\eta^{\mu\nu} - k_{1}^{\nu}k_{2}^{\mu}\right],
\end{aligned}
\end{equation}
\begin{equation}
\begin{aligned}
&V(h, \gamma_{\mu}(k_{1}), Z_{\nu}(k_{2})) =
iC_{\gamma Zh}\left[(k_{1}k_{2})\eta^{\mu\nu} - k_{1}^{\nu}k_{2}^{\mu}\right]\\
&=  -i\dfrac{\alpha}{2\pi\upsilon_{0}}\left(2\gamma b \left(\frac{b_{2}}{tan \theta_{W}} - b_{Y}tan \theta_{W}\right)- (d + \gamma b) (A_{F} + A_{W}) \right) \left[(k_{1}k_{2})\eta^{\mu\nu} - k_{1}^{\nu}k_{2}^{\mu}\right],
\end{aligned}
\end{equation}
\begin{equation}
\begin{aligned}
&V(\phi, \gamma_{\mu}(k_{1}), Z_{\nu}(k_{2})) =
iC_{\gamma Z \phi}\left[(k_{1}k_{2})\eta^{\mu\nu} - k_{1}^{\nu}k_{2}^{\mu}\right]\\
&=  -i\dfrac{\alpha}{2\pi\upsilon_{0}}\left(2\gamma a \left(\frac{b_{2}}{tan \theta_{W}} - b_{Y}tan \theta_{W}\right)- (c + \gamma a) (A_{F} + A_{W}) \right) \left[(k_{1}k_{2})\eta^{\mu\nu} - k_{1}^{\nu}k_{2}^{\mu}\right],
\end{aligned}
\end{equation}
where $b_{3} = 7, b_{2}= 19/6, b_{Y} = - 41/6$ are the $\mathrm{SU}(2)_{L} \otimes
 \mathrm{U}(1)_{Y}$ $\beta$-function coefficients in the SM.\\
The auxiliary functions of the $h$ and $\phi$ are given by
\begin{align}
&F_{1/2}(\tau) = -2 \tau[1 + (1-\tau) f(\tau)],\\
&F_1(\tau) = 2 + 3\tau + 3\tau(2-\tau) f(\tau),
\end{align}
with
\begin{align}
&f(\tau ) = \left(\sin ^{-1} \frac{1}{\sqrt{\tau } } \right)^{2} \, \, \, (for\, \, \, \tau >1),\\
&f(\tau ) = -\frac{1}{4} \left(\ln \frac{\eta _{+} }{\eta _{-} } - i\pi \right)^{2} \, \, \, (for\, \, \, \tau <1),\\
&\eta _{\pm} = 1\pm \sqrt{1 - \tau } ,\, \, \, \tau _{i} = \left(\frac{2m_{i} }{m_{s} } \right)^{2} .
\end{align}
 $m_{i}$ is the mass of the internal loop particle (including quarks, leptons and W boson), $m_{s}$  is the mass of the scalar state ($h$ or $\phi$). Here, $\tau _{f} = \left(\frac{2m_{f} }{m_{s} } \right)^{2},   \tau _{W} = \left(\frac{2m_{W} }{m_{s} } \right)^{2}$ denote the squares of fermion and W gauge boson mass ratios, respectively.\\
 \newpage
  \begin{center} 
 { APPENDIX B: Feynman diagrams for the considered process}\\
 \end{center}
\end{document}